\documentclass[ aip, PoF,
amsmath,amssymb,
reprint,%
]{revtex4-2}

\usepackage{dsfont}
\usepackage{mathtext}
\usepackage[utf8]{inputenc}

\usepackage{mathtools}
\usepackage[usenames,dvipsnames]{xcolor}
\usepackage{amsmath}
\usepackage{amssymb,mathrsfs}
\usepackage{graphicx}
\usepackage{epstopdf}
\usepackage[colorlinks=true]{hyperref}

\usepackage{mathtext}
\usepackage{subcaption}

\usepackage{etoolbox} % for \appto
\usepackage{lipsum} % for mock text
\usepackage[capitalize]{cleveref}

\makeatletter
\appto{\appendix}{%
	\@ifstar{\def\theequation@prefix{A.}}%
	{}%
}
\makeatother

\newcommand{\om}{\omega}

\newcommand{\prt}{\partial}

\newcommand{\RNumb}[1]{\uppercase\expandafter{\romannumeral #1\relax}}

\begin{document}
\title{Propagation of narrow and fast solitons through dispersive shock waves in hydrodynamics of simple waves}

\author{D.~V.~Shaykin}
\affiliation{Russian University of Transport (RUT-MIIT), Obrazcova st. 9, Moscow, Russia}
\affiliation{Skolkovo Institute of Science and Technology, Skolkovo, Moscow, 143026, Russia}
\affiliation{Moscow Institute of Physics and Technology, Institutsky lane 9, Dolgoprudny, Moscow region, 141700, Russia}

\begin{abstract}

We study the propagation of narrow solitons through various profiles of dispersive shock waves (DSW) for the generalized Korteweg-de Vries equation. We consider situations in which the soliton passes through the DSW region quickly enough and does not get trapped in it. The idea is to consider the motion of such solitons through DSW as motion along some smooth effective profile. In the case of KdV and modified KdV, based on the law of conservation of momentum and the equation of motion, this idea is proven rigorously; for other cases of generalized KdV, we take this as a natural generalization. In specific cases of self-similar decays for KdV and modified KdV, a method for selecting an effective field is demonstrated. For the case of generalized KdV, a hypothesis is proposed for selecting an effective field for any wave pulse that is not very large compared to the soliton. All proposed suggestions are numerically tested and demonstrate a high accuracy of reliability.

\end{abstract}

\maketitle

\section{Introduction}

This article is a natural continuation of a series of articles in which a new approach to studying the dynamics of solitons, linear wave packets, and related problems was developed\cite{{El-05},{El-08},{El-19},{kamch-19},{kamch-20},{kamch-shayk-21},{kamch-shayk-23-1},{kamch-shayk-23-2},{kamch-23},{kamch-shayk-24}}. This method allows avoiding the use of quite complex mathematical theories, such as the inverse scattering problem method, the Riemann-Hilbert problem, and others, which are applicable only to integrable equations (with permissible perturbations). In the generalized KdV (gKdV) equation of interest to us, only cases KdV and modified KdV (mKdV) are integrable. 

A history of this method mainly takes its beginnings with El's work\cite{El-05} where was performed a new analysis of dispersive shock waves (DSW) in case of generalized Korteveg-de Vries equation
\begin{equation}\label{int1}
u_t+V(u)u_x+u_{xxx}=0.
\end{equation}
There was under consideration the well known Riemann problem of the breakup of the initial discontinuity with an appeared dispersive shock wave . An applied here standard technique of Whitham's modulation equations\cite{{Whitham-74},{Whitham-65}} gave a new result: in the limit cases of the modulation system, usually named soliton edge and small amplitude edge, there was revealed a new type of differential dependency between parameters of the DSW. In particular, for the small amplitude edge we can write an expression connecting wave number $k = 2\pi/\lambda$ of linear wave at the corresponding front of the DSW with the constant value of the background field $u$
\begin{equation}\label{int2}
\frac{\prt k}{\prt u} = \frac{\prt \om(u,k)/\prt u}{V(u)- \prt \om(u,k)/\prt k},
\end{equation} 
where $\om(u,k) = V(u)k-k^3$ being the dispersion law for the linear waves obtained from the linearised Eq.\eqref{int1} (a similar dependence was identified on the soliton edge). This made it possible to repeat\cite{El-08} already known results\cite{Karpman-67} of parameters of solitons generated from the initial pulse but without using the integrability of the KdV.

After some time, a new derivation\cite{{kamch-19},{kamch-20}} of this relation was provided using an optical-mechanical analogy, which allowed to investigate the dynamics of the fronts of DSWs on inhomogeneous and non-stationary waves in noninegrable equations. Fundamental results on the problem of soliton formation from an initial impulse for simple waves, that is, waves propagating in one direction, were also generalized. 

At the same time, studies\cite{{El-19},{kamch-shayk-21}} of the motion of a separate linear wave packet were initiated. One of the key roles is played by the optical-mechanical analogy for the linear wave packet (see, for example, Ref.\cite{LL-TP}) based on separation of scales. Namely, there are different scales: characteristic size of large background $L$ and the high-frequency linear wave packet with wave number $k = 2\pi/ \lambda$ and longitudinal size $l$  such as
\begin{equation}\label{int3}
\lambda \ll l \ll L.
\end{equation} 
Owing to this remark, one can consider the linear wave packet as a mechanical particle with a Hamiltonian $\omega (u,k)$ and a momentum $k$, that means dynamic equations are
\begin{equation}\label{int4}
\frac{dx}{dt} = \frac{\prt \om(u,k)}{\prt k}, \qquad \frac{dk}{dt} = -\frac{\prt \om(u,k)}{\prt x}.
\end{equation} 
Such a way made it possible to solve the problem of the propagation of the linear wave packet along any large background in the case of unidirectional motion\cite{kamch-shayk-21}. By solving equations \eqref{int1} and \eqref{int4} together, one can obtain dependence \eqref{int2} without complex analysis of characteristics in the Whitham theory. A trivial integration leads to the dependency between the carrier wave number and the background field
\begin{equation}\label{int7}
k^2 = \frac{2}{3}V_0(u)+q,
\end{equation}
where $q$ is just a constant of integration. But it turns out that $q$ plays the key role in the dynamics of the packet because it, as well as the first integral, carries information about the packet. So, the joint evolution of the wave packet and the background can be described by system
\begin{equation}\label{int8}
u_t+V_0(u)u_x = 0, \qquad \frac{dx}{dt} = -V_0(u)-3q,
\end{equation}  
where the first one is well-known Hopf's equation and the second is an ordinary differential equation. 

The next stage of development of the approach was generalization to physical systems described by several independent functions; for example, in hydrodynamics they are always used as density $\rho$ and flow velocity $u$. The optical-mechanical analogy made it possible to solve the same problem in such cases too. It was revealed that one can obtain the exact solution for the wave number $k$ only if the equation is integrable\cite{kamch-shayk-24}, in other cases the method can permit an approximate solution, where constant $q$ plays an important role. Apparently, the presence of an exact solution can serve as a new very simple test of the integrability of nonlinear equations. As  well, there was revealed the physical meaning of well-known operators $\mathcal{A}$ and $\mathcal{B}$ in the Lax pair\cite{{kamch-23},{kamch-shayk-24}}.  

The second important role is played by Stokes's remark, pointed out in a letter\cite{Stokes} from Stokes to Lamb, that allows linking the solution of the linear problem with the soliton motion problem. Recently, this made it possible to generalize\cite{kamch-23} the problem of soliton characteristics\cite{{Karpman-67},{El-08}} after the breakdown of an arbitrary profile to hydrodynamics $\{\rho,u\}$. 

The remark states since the tail of a soliton has a small amplitude, then it can be governed by a linearized equation, and moreover the tail drops exponentially. As a consequence, there is an asymptotic correspondence with an ordinary linear wave:
\begin{equation}\label{int9}
 e^{-k_s(x-V_st)}\quad \text{and} \quad  e^{i(kx-\om t)} \qquad  
\end{equation}
that leads to the correspondence
\begin{equation}\label{int10}
k_s = ik \qquad V_s = \frac{\om(k_s)}{k_s}.
\end{equation}
where $k$ and $\omega$ are the same that were for the linear waves. As a result, the dynamics of the solitons can be described by means of only linear problem that has just been solved. So, for the soliton dynamics we have
\begin{equation}\label{int11}
k_s^2 = -\frac{2}{3}V_0(u)+q, \quad \frac{dx}{dt} = \frac{1}{3}V_0(u)+q, \quad u_t + V_0(u)u_x = 0.
\end{equation} 
But we have to take into account it works in the limit of high frequency wave packet, that is, $k \gg 1$ and as consequence $k_s \gg 1$. Therefore, such a way we can describe only a sufficiently narrow soliton, namely, value $u$ along its length should hardly change.

Thus, as one can see, this new approach is very powerful and fundamental. It is based on the common observation and it is not connected with the integrability of the equations. It does not require the shape of waves, background or solitons but gives the most simple and powerful dynamics equations  suitable for the most general initial conditions.

So, in the current paper we want to apply this technique to an actively developing problem of propagation of solitons through the dispersive shock waves (see Ref.\cite{{2018},{2021}}).  Our attention was drawn to article Ref.\cite{Ablow-23} (see many useful references therein), where different approaches to this problem were discussed, considering the Riemann problem for the Korteweg-de Vries equation. In the mentioned work, an attempt was made in the first section to work with the known integral conservation laws in the KdV theory; here, we will show that this work can lead to the generalization of equations \eqref{int11} for an effective field that is not required to obey the Hopf equation (the third equation in \eqref{int11}). Here we will demonstrate that by means of the discussed method, we can extend investigation on other initial conditions; moreover, we will give recipes of solutions to the problem for several typical situations. But the main point here is our study will not be expanded on  cases that were called "trapped soliton" in Ref.\cite{Ablow-23}, hence the soliton under our consideration, even during its propagations through the DSWs, will have the shape of the solitons but not a "dislocation", this is essential. We will also consider sufficiently fast solitons; some details of an opposite problem can be found in Ref.\cite{Ablow-23}.

The main idea is as follows: DSW is just a train of waves; hence, If the soliton passes through the oscillations of the DSW quickly, then the interaction with each DSW oscillation can be replaced with a single ``effective'' field (see Kapitza's pendulum\cite{LL-M}). If it can be performed in this way, we can use the mentioned above Eq. \eqref{int11} and the problem is done. But on this stage we face the problem of choosing the appropriate effective field that will be discussed further.

\section{Korteweg-de Vries equation}

First of all, let us investigate the propagation of solitons through DSWs in the example of the well-known and the most simple case of KdV. To emphasize that we work with narrow solitons (we want to escape a complicated problem of interaction between solitons and background that can lead to some radiation etc), we put here the small dispersion parameter $\varepsilon$
\begin{equation}\label{kdv1}
u_t +6uu_x+\varepsilon^2u_{xxx} = 0,
\end{equation}
where $\varepsilon\ll 1$. As is known, such an equation has the solution in the form "background+soliton" 
\begin{equation}\label{kdv2}
u(x,t) = u_b+\frac{V-6u_b}{\cosh^2(\sqrt{(V-6u_b)/2\varepsilon^2}(x-Vt))},
\end{equation}
where $u_b$ describes the background, and one can see from this equation the dependency of soliton parameters from the $u_b$ $(V>6u_b!)$. We will suggest there is an absence of the influence of the soliton propagation on the background evolution, so it leads to the well-known Hopf's equation 
\begin{equation}\label{kdv2.1}
u_{b_t}+6u_b u_{b_x} = 0
\end{equation}
managing the evolution of smooth waves.

Generally speaking, we are interested in the propagation of the solitons through fast-oscillating regions where Eq.\eqref{kdv2.1} doesn't work. So, let us assume that we can describe such a propagation as a motion along an  "effective" background. Obviously, the effective background no longer obeys the Hopf equation, but we will assume that the soliton parameters still depend the same way as in Eq.\eqref{kdv2} on the effective background, so it leads to
\begin{equation}\label{kdv3}
u(x,t) = w(x,t)+\frac{V-6w}{\cosh^2(\sqrt{(V-6w)/2\varepsilon^2}(x-Vt))},
\end{equation}
where we use the notation $w$ for the effective background instead of the previously used $u_b$ $(V>6w!)$. In short, Eq. \eqref{kdv3} can be written as
\begin{equation}\label{kdv4}
u(x,t) = w(x,t)+ v(x,t;w).
\end{equation}
As soon as we look for such a type of solution, we should substitute it into Eq. \eqref{kdv1}, so it brings us to
\begin{equation}\label{kdv5}
v_t + 6(wv)_x + 6vv_x +\varepsilon^2v_{xxx} = -F[w(x,t)],
\end{equation}
where following Ref.\cite{Ablow-23} we introduce the notation
\begin{equation}\label{kdv6}
F[w(x,t)] = w_t + 3(w^2)_x+ \varepsilon^2w_{xxx}.
\end{equation}
Here one can notice  if $w$ coincides with the natural background $u_b$, then $F[w] = 0$ in the same order of $\varepsilon$ in which we obtain \eqref{kdv2.1}. Further, we introduce two ordinary dynamic equations for the soliton: change of momentum and center of mass. The first one is
\begin{equation}\label{kdv7}
\begin{split}
\frac{d}{dt} \int_\mathbb{R}v^2dx = &-2 \int_\mathbb{R} vF[w]dx - 12 \int_\mathbb{R} v(wv)dx -\\ &-2\varepsilon^2 \int_\mathbb{R} vv_{xxx}dx  - 12 \int_\mathbb{R} v^2v_xdx
\end{split}
\end{equation} 
and the evolution of the soliton center of mass
\begin{equation}\label{kdv8}
\begin{split}
\frac{d}{dt} \bigg( \frac{\int x v^2dx}{\int v^2dx} \bigg) = \frac{\frac{d}{dt}\int xv^2dx}{\int v^2dx} - \frac{(\int xv^2dx)\frac{d}{dt}\int v^2dx}{(\int v^2dx)^2},
\end{split}
\end{equation} 
 where for brevity we omit the limits of integrations implying they are the same as in the previous formula. Eq. \eqref{kdv7} obtained by multiplying Eq.\eqref{kdv5} by $v$ and integrating over $\mathbb{R}$.
 
 First, let's deal with the evolution of the momentum, and after that we apply the same manipulations to the longer expression of the evolution of the mass center. So, as it was discussed above, we consider the medium with a weak dispersion ($\varepsilon \ll 1$), which means the soliton in Eq. \eqref{kdv3} becomes very thin. Therefore, it is convenient to deal with the soliton $v(x,t;w)$ like with the delta function, implying that other functions change significantly slower. In particular, in the limit $\varepsilon \rightarrow 0$ the integrals from Eq.\eqref{kdv7} can be calculated as the integrals with the delta functions (see Appendix A). Then we obtain a considerably simplified  expression 
\begin{equation}\label{kdv9}
\frac{dk^2}{dt} = -4w_xk^2 - 4\overline{F}[w],
\end{equation}
where $\overline{F}[w]$ is the notation of relation \eqref{kdv6} in the limit transition $\varepsilon \rightarrow 0$, that is, $\overline{F}[w] = w_t + 6ww_x$.

Similarly, multiplying formula \eqref{kdv5} by $x \cdot v$ and by $v$, for the center of mass we come to a long expression
\begin{equation}\label{kdv10}
\begin{split}
\frac{dx_s}{dt}= &-\frac{ \int  2xv\{F[w] + 6(wv)_x + 6vv_x + \varepsilon^2v_{xxx} \}dx}{\int v^2dx} \\
&+ x_s \frac{\int \{6w_xv^2 + 2vF[w]\}dx}{\int v^2dx},
\end{split}
\end{equation}
where in the second term we have used the definition for the center of mass entered in Eq.\eqref{kdv8} and  already familiar substitution $\frac{d}{dt}\int v^2dx$. Taking into account the well-known manipulations with the delta function (see Appendix A), we arrive at a simplified shape of Eq.\eqref{kdv10}:
\begin{equation}\label{kdv11}
\frac{dx_s}{dt} = 6w+k^2.
\end{equation}
Thus, we obtain the system of two equations
\begin{equation}\label{kdv12}
\frac{dk^2}{dt} = -4w_xk^2 - 4(w_t+6ww_x),\quad
\frac{dx_s}{dt} = 6w+k^2,
\end{equation}

Let's assume that parameters of the soliton depend on the effective background, that is, $k^2 = k^2(w)$, hence
\begin{equation}\label{kdv13}
\frac{dk^2}{dt} = \frac{dk^2}{dw}\frac{dw}{dt}\bigg|_{x=x_s(t)} = \frac{dk^2}{dw}\big( w_t+\frac{dx_s}{dt}w_x \big).
\end{equation}
The substitution of system \eqref{kdv12} in Eq. \eqref{kdv13} leads to 
\begin{equation}\label{kdv14}
\frac{dk^2}{dt} = -\frac{1}{4}\frac{dk^2}{dw}\frac{dk^2}{dt}
\end{equation} 
that have the solution 
\begin{equation}\label{kdv15}
k^2 = -4w + q
\end{equation}
with constant $q$, this yields
\begin{equation}\label{kdv16}
\frac{dx_s}{dt} = 2w + q.
\end{equation}
These formulas coincide with Eq.\eqref{int11} for case $V_0 = 6u$, but now we have not used $w$ as a solution of the Hopf equation. It means such formulas can be extended to a case of fast-oscillating regions, where it is convenient  in a sense to consider the effective background. Therefore, all we have to do is to choose the effective field for the soliton and substitute it into Eq. \eqref{kdv16}

\subsection*{ Propagation of KdV solitons through DSWs arising after self-similar wave breaking}
So, to apply these lengthy arguments, for the beginning we propose to consider the Riemann problem
\begin{equation}\label{kdv17}
u_0(x) = 
\begin{cases}
1 \quad x<0 \\
0 \quad x>0
\end{cases}
\end{equation}
As was studied in Ref.\cite{gur-pit-73} this profile transforms to DSW, where the left and the right edges move according to
 \begin{equation}\label{kdv18}
x_r = 4t, \qquad x_l = -6t.
\end{equation}
Due to Galileo’s invariance, this case generalizes other steps-like forms. So, in Ref.\cite{Ablow-23} it was suggested to take the bottom envelope of the DSW as a mean field for the soliton propagation, it almost coincided with the straight line between the left and the right edges. Here is the idea: if we consider the DSW as a train of soliton-like oscillations, then the external soliton will shift after each interaction with solitons in the train. In the case of a small amplitude train in comparison with the external soliton, we can suggest that such a total shift will be negligible and the bottom envelope is a good suggestion for the mean-field. This dependence on the amplitude of the soliton (and hence its velocity, too) has been confirmed numerically.

Apparently, it is a complicated task to propose another effective field except a straight line
\begin{equation}\label{kdv19}
w_{\textit{{\tiny DSW}}} = 0.4-0.1\frac{x}{t},
\end{equation}
between \eqref{kdv18} points, but what should one do in another initial distribution? We want to demonstrate a new heuristic principle that gives excellent agreement with numerical calculation. 

So, any initial distribution at breaking time can be decomposed into a Teylor series near the breaking point, hence an important problem here is to solve the propagation of solitons through the power-like initial distributions
\begin{equation}\label{kdv20}
u_0(x) = 
\begin{cases}
(-x)^{1/n} &\quad x \leq 0 \\
0 		   &\quad x>0
\end{cases}
\end{equation}
We propose a new principle: in case \eqref{kdv20} we cannot prefer any other effective profile other than a power one with power $n$, namely, $w_{\textit{{\tiny DSW}}}$ is like $\sqrt[n]{-x}$. Obviously, during an evolution the DSW expands that means we need to shift and scale this effective background. So, we want to build 
\begin{equation}\label{kdv21}
w_{\textit{{\tiny DSW}}}(x,t;n) = \sqrt[n]{\frac{a_n(t)-x}{b_n(t)}}
\end{equation}
It is great that we need only two unknown parameters $a_n, b_n$ while we have two conditions at the edges of the DSW. From the  Ref.\cite{KamchN}, we know the law of the edges motion of such a self-similar wave breaking:
\begin{equation}\label{kdv22}
x_l(t) = -\bigg(1-\frac{1}{n}\bigg) \bigg(2-\frac{1}{n} \bigg)^\frac{1}{n-1}\bigg(6t\bigg)^{\frac{n}{n-1}},
\end{equation}
\begin{equation}\label{kdv23}
x_r(t) = \bigg(1-\frac{1}{n}\bigg) \bigg(\frac{\Gamma (2n+2)}{n[\Gamma (n+1)]^2} \bigg)^\frac{1}{n-1} t^{\frac{n}{n-1}},
\end{equation}
where two of the most typical situations $n = \{2,3\}$ as well were explored in Ref.\cite{{gur-pit-74},{pot-88},{dub-93}}:
\begin{equation}\label{kdv24}
\begin{split}
&n=2 \qquad\quad x_l = -27t^2,\quad x_r = 15t^2
\\
&n=3 \qquad \quad x_l = -4\sqrt{10t^3}, \quad x_r = \frac{4}{3}\sqrt{\frac{35t^3}{3}}
\end{split}
\end{equation}
\begin{figure}[t]
\begin{center}
	\includegraphics[width = 7cm,height = 7cm]{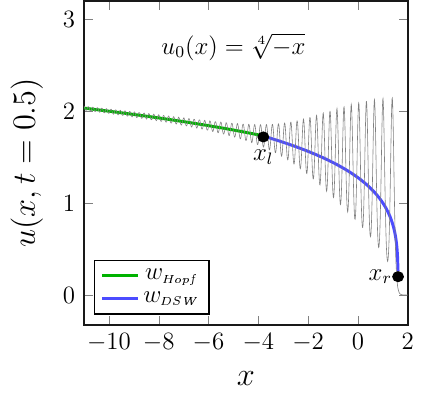}
\caption{Image of an dispersive shock wave arising after breaking of $u_0 = \sqrt[4]{-x}$. Blue line indicates the effective field obtained by supposed in this section method. Green line indicates the effective field coincides with the Hopf solution outside the DSW.}
\label{fig1}
\end{center}
\end{figure}
It is important to emphasize that in our reasoning we did not rely on the interaction between DSW oscillations and external solitons, that is, phase shifts were completely bypassed in our principle; we looked only at the analytical form of the initial distribution. It means that such a theory can work with any amplitude and velocities of solitons if only solitons are sufficiently narrow to use the local formulas \eqref{kdv16},\eqref{kdv17}.

 Let's demonstrate how we can build the effective background $w_{\textit{{\tiny DSW}}}$ for $n=2$ in Eq. \eqref{kdv21}. If we consider the right edge of the DSW, we find out there $w(x_r,t) = 0$, hence $a_2(t) = x_r(t)$ and it works for any $n$. Then we look at the left edge and match $w_{\textit{{\tiny DSW}}}$ with Hopf's solution outside the DSW region. It leads to
\begin{equation}\label{kdv25}
\begin{cases}
x_l -6ut = -u^2,
\\
 \sqrt{\frac{x_r-x_l}{b_2}} = u 
\end{cases}
\quad \longrightarrow \quad b_2 =\frac{1}{9}\sqrt{\frac{69}{2}}.  
\end{equation} 
Finally it leads to 
\begin{equation}\label{kdv26}
w_{\textit{{\tiny DSW}}}(x,t;2) = 9\sqrt{\frac{15t^2-2x}{69}}.
\end{equation}
It is interesting that such a profile also coincides with the bottom envelope of the DSW, but this observation does not work for all $n$ except the step-like and $n=2$. 
\begin{figure}[t]
\begin{center}
	\includegraphics[width = 7cm,height = 7cm]{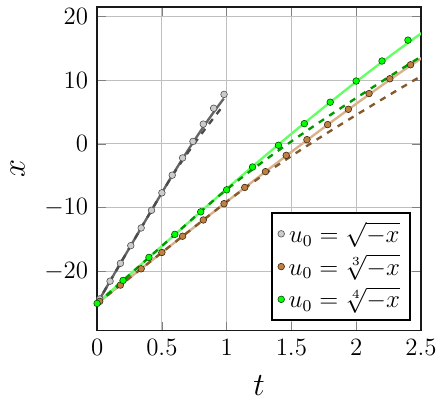}
\caption{Paths $x(t)$ of the solitons for three situations $\{x_0,V_0, u_0\}$: grey colour is $\{-25,35,\sqrt{-x}\}$, green colour is $\{-25,18,\sqrt[3]{-x}\}$, brown colour is $\{-25,16,\sqrt[4]{-x}\}$. The circle indicate the data obtained by numerical calculation of Eq.\eqref{kdv1} with $\varepsilon = 0.1$, the solid lines are the analytical predictions of the paths along the effective field \eqref{kdv21}, the dashed line are the paths along the effective field got from the straight line between edges of the DSW. The curves end where the solitons exit the DSW. 
}
\label{fig2}
\end{center}
\end{figure}
To demonstrate the effectiveness of our effective field \eqref{kdv21}, we tested it for $n= 2,3,4$. A typical image of the $w_{\textit{{\tiny DSW}}}(x,t;n)$ is depicted in the Fig.\ref{fig1} by a blue line. In contrast to the proposed theory, one can use a straight  line between $x_l,x_r$ which is also depicted in Fig.\ref{fig1}, but it turns out that such an approximation can be inaccurate, as we demonstrate in Fig.\ref{fig2} where there is a significant difference in the paths of the solitons between these two approaches. We will discuss in the third section situations when such a linear approximation is justified.

\section{modified Korteweg-de Vries}

 Now we will consider another equation which permits us to make the same manipulation, but for a dark soliton. So, let's turn to the modified Korteveg-de Vries equation 
\begin{equation}\label{mkdv1}
u_t -6u^2u_x+\varepsilon^2u_{xxx} = 0.
\end{equation} 
  Here the standard technique allows to obtain the form of a dark soliton on the background wave:
\begin{equation}\label{mkdv2}
u = u_b - \frac{V+6u_b^2}{2u_b+\sqrt{-V-2u_b^2}\cosh{\sqrt{(V+6u_b^2)/\varepsilon^2}(x-Vt)}},
\end{equation}
where, as one can see, the dependence of soliton parameters on the external impulse is more complex than it was in the case of KdV. So, as it was earlier we look for a solution
\begin{equation}\label{mkdv3}
u(x,t) = w(x,t)+v(x,t;w)
\end{equation}
where $w$ being an effective background and $v$ is the soliton mKdV. Substitution of Eq. \eqref{mkdv3} into Eq. \eqref{mkdv1} leads to
\begin{equation}\label{mkdv4}
v_t-6v^2v_x-6(v^2w)_x-6(vw^2)_x+\varepsilon^2v_{xxx} = -F[w],
\end{equation}
here $F[w]$ also denotes the mKdV operator for the effective field: $F[w] = w_t-6w^2w_x+\varepsilon^2w_{xxx}$.

Doing similar calculations as it has been made for the KdV (Appendix A) we arrive at the laws for the changing of the momentum and the center of mass of the soliton mKdV, there are
\begin{equation}\label{mkdv5}
\frac{d}{dt}\int v^2  = -2\overline{F}[w]\cdot\int v + 6 (w^2)_x\cdot \int v^2 + 4w_x\cdot \int v^3, 
\end{equation} 
\begin{equation}\label{mkdv6}
\frac{dx_s}{dt} =-\frac{1}{\int v^2 }\bigg[  6w^2\cdot \int v^2 + 8 w\cdot\int v^3 + 3\cdot\int v^4 + 3 \varepsilon^2 \cdot\int (v_x)^2  \bigg], 
\end{equation}
where for brevity $dx$ is omitted. In Appendix B we accurately reduce these relations to the simple form
\begin{equation}\label{mkdv7}
\frac{dk^2}{dw^2} =4,  \quad \frac{dx_s}{dt} = -2w^2-q.
\end{equation}
It again leads to the familiar formulas \eqref{int11}
\begin{equation}\label{mkdv8}
k^2 = 4w^2-q,  \qquad \frac{dx_s}{dt} = V =  -2w^2-q.
\end{equation}
In the case of the dark soliton mKdV \eqref{mkdv2} we have some restriction on parameters of the soliton, simple analysis gives
\begin{equation}\label{mkdv9}
-6w^2< V < -2w^2, \qquad 0 < q < 4w^2, \qquad 0<k^2<4w^2. 
\end{equation}

\subsection*{Propagation of mKdV solitons through DSWs arising after self-similar wave breaking}
 
Here we will develop a theory for the selection of effective fields for the mKdV in a particular case of breaking of the following initial profiles
\begin{equation}\label{mkdv10}
u_0(x) = 
\begin{cases}
\sqrt{(x)^{1/n}+u_b^2} &\quad x\geq 0,
\\
u_b &\quad x<0,
\end{cases}
\end{equation}
\begin{figure}[t]
\begin{center}
	\includegraphics[width = 7cm,height = 7cm]{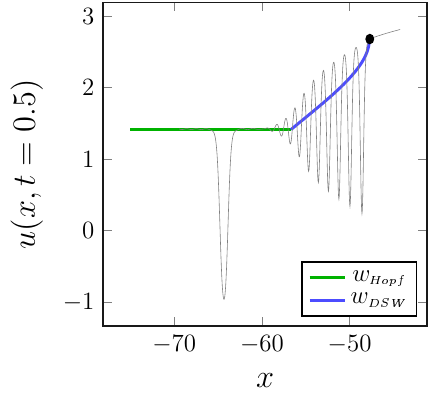}
\caption{The figure shows the numerical calculation of Eq.\eqref{mkdv1} with $\varepsilon=0.5$ for time $t=0.5$ for the initial profile \eqref{mkdv10} and soliton \eqref{mkdv2} located at point $x_0=-30$ at the initial moment in time. The choice of effective field for equation of motion \eqref{mkdv8} is indicated in green and blue. The blue line corresponds to effective field \eqref{mkdv16}, and the green line to the constant field $u_b=\sqrt{2}$.
}
\label{fig2.1}
\end{center}
\end{figure}
where background value $u_b$ cannot be omitted since Eq. \eqref{mkdv1} is not Galilean invariant, hence there isn't convenient variables replacement allowing to get rid of the $u_b$ in contrast to Eq. \eqref{kdv1}. The wave breaking of such profiles in Eq. \eqref{mkdv10} as well leads to the formation of the DSW. Looking for the edges of DSW for Eq. \eqref{mkdv1} with the initial condition \eqref{mkdv10} reduces to the following equation
\begin{equation}\label{mkdv11}
x+6u^2t = \overline{x_0}(u^2),
\end{equation} 
that as well was studied in Ref.\cite{Kamch-Kon-04}(section 3). There was used the generalized hodograph method (see in detail Ref. \cite{{Tsarev},{Gurevich}}). If we work in terms of $u^2$, then the problem reduces\cite{Kamch-Kon-04} to the KdV case, that is why we can treat such a problem by means of the proposed earlier principle and at the same time, test it.

So, for example, applying the hodograph method to the initial profile
\begin{equation}\label{mkdv12}
u_0^2(x) = 
\begin{cases}
(x)^{1/2}+u_b^2 &\quad x\geq 0,
\\
u_b^2 &\quad x<0,
\end{cases}
\end{equation}
we can obtain the following DSW edges:
\begin{equation}\label{mkdv13}
x_l = -6(u_b)^2t-\frac{27}{2}t^2 \qquad x_r = -6(u_b)^2 - 5t^2.
\end{equation}
\begin{figure}[t]
\begin{center}
	\includegraphics[width = 7cm,height = 7cm]{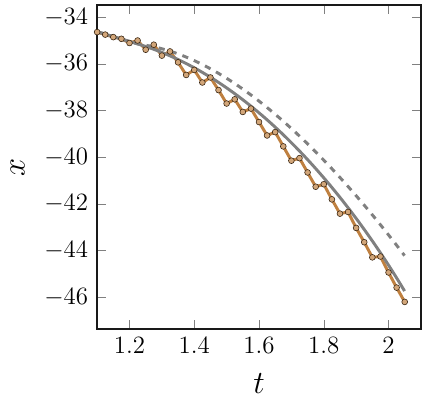}
	\caption{The dependence of the soliton's path on time for Eq.\eqref{mkdv1} with $\varepsilon = 0.5$. The gray lines represent analytical predictions calculated using the formula \eqref{mkdv7}; the dashed line corresponds to the straight effective field between edges \eqref{mkdv13} of the DSW obtained from the initial condition \eqref{mkdv12}; the solid gray line corresponds to the effective field \eqref{mkdv16}.  The brown circles connected by a line correspond to the numerical calculation for the soliton \eqref{mkdv2}, initially positioned on a background of $u_b=\sqrt{2}$ with an initial coordinate of $x_0=-30$ and having a speed of $V_0=-4.2$.
}
\label{fig2.2}
\end{center}
\end{figure}
It can be noted that we can obtain an analogous laws for any $n$ in Eq. \eqref{mkdv10}. To build an effective background for the soliton motion, we should take into account the main principle that we need to use the same order as it was in the initial profile. But here we have the dark solitons train in DSW, so we need to adapt our main principle, and we should look for the effective background as
\begin{equation}\label{mkdv14}
w_{\textit{{\tiny DSW}}}^2(x,t;n=2) = u_{\textit{{\tiny Hopf}}}^2(x_r)-\sqrt{\frac{a-x}{b}},
\end{equation}
where $u_{\textit{{\tiny Hopf}}}$ is a solution of the Hopf equation 
\begin{equation}\label{mkdv15}
x+6u^2t=(u^2-u_b^2)^2
\end{equation}
The ``-'' sign before the root in Eq.\eqref{mkdv14} becomes obvious when looking at the form of the DSW in Fig.\ref{fig2.1}, where the ``inverted'' version of Fig.\ref{fig1} is presented. 

To obtain the heuristic form \eqref{mkdv14}, we decide to put the left edge of the effective curve $\sqrt{-x}$ at the constant value $u_b$ and the right edge at the Hopf solution $u_{\textit{{\tiny Hopf}}}^2(x_r)$ associated with the edges of the DSW, it leads to
\begin{equation}\label{mkdv16}
w_{\textit{{\tiny DSW}}}^2(x,t;2) = u_b^2+5t-\sqrt{\frac{-50}{17}\big( 6u_b^2+5t^2+x \big)}
\end{equation}

If we are interested in the propagation of solitons through high-frequency DSW, we obtain the same picture as was in Fig.\ref{fig2} demonstrating the great advance of proposed power-like effective fields over trivial straight lines. So, to present some interesting results, we will consider not sufficient high-frequency DSW (in comparison with the length of the external solitons). Generally speaking, our theory is based on the law of motion of a soliton along a non-uniform profile, in which, as shown in Ref.\cite{kamch-shayk-23-2}, the assumption of the soliton's narrowness is crucial. In the current article, we have transferred this requirement to the soliton's narrowness compared to the background oscillation. In this regard, the idea arises to find out what will happen with a sufficiently wide soliton and a weakly oscillating background. Let us consider the situation depicted in Fig.\ref{fig2.1}. Fig.\ref{fig2.2} presents a comparison of the numerical calculation with the analytical prediction, as if it were possible to introduce an effective background. As we can see, the soliton returns to the line we predicted after each interaction with the background oscillation. Thus, even in such a specific case, the effective profile we proposed gives good agreement with the numerical calculation.

\section{Generalized Korteweg-de Vries equation} 

Now we shall try to extend the proposed approach to a generalized equation
\begin{equation}\label{gkdv1}
u_t +6u^pu_x+\varepsilon^2u_{xxx} = 0.
\end{equation}
and to the case of not only self-similar solutions. For definition, further in this section we will assume wave-breaking into the resting part of medium and wave-breaking occurs at the front of the initial pulse. For the soliton, despite the fact that we cannot obtain a solution "background+soliton" and apply the integral technique developed earlier, we will assume there is the same result
\begin{equation}\label{gkdv2}
\frac{dx}{dt} = \frac{1}{3}V_0(w)+q.
\end{equation}

Usually, in applications we have an initial breaking pulse of limited size, and then, as was demonstrated in Ref.\cite{kamch-19} we know a law of motion of a small amplitude edge as a motion of the linear wave packet in the form of Eq.\eqref{int8}
\begin{figure}[t]
\begin{center}
	\includegraphics[width = 7cm,height = 7cm]{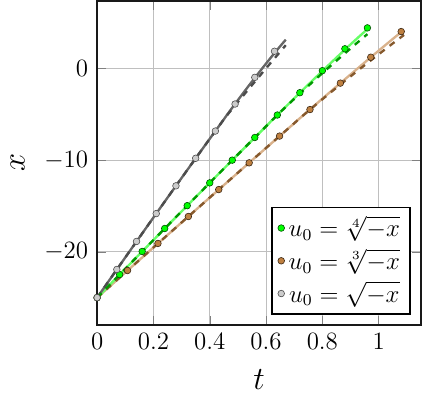}
\caption{Paths $x(t)$ of the solitons for three situations $\{x_0,V_0, u_0\}$: grey colour is $\{-25,7+6\sqrt{25},\sqrt{-x}\}$, green colour is $\{-25,7+6\sqrt[3]{25},\sqrt[3]{-x}\}$, brown colour is $\{-25,17+\sqrt[4]{25},\sqrt[4]{-x}\}$ with identical initial amplitude $7$. The circles indicate the data obtained by numerical calculation of Eq.\eqref{kdv1} with $\varepsilon = 0.1$, the solid lines are the analytical predictions of the paths along the effective field \eqref{kdv21},  the dashed lines are the paths along the effective fields as straight lines  between edges of the DSW. The end of the lines corresponds to the soliton exiting the DSW. 
}
\label{fig4}
\end{center}
\end{figure}
\begin{equation}\label{gkdv3}
\frac{dx_l}{dt} = -V_0(u)-3q = -V_0(u),
\end{equation} 
where $q$ we should take as $0$ due to the initial condition at the wave-breaking point, namely, $k=0$ at $u=0$. Likewise, it was demonstrated that an asymptotic law of motion for a soliton edge is obtained by using Eq.\eqref{int11} 
\begin{equation}\label{gkdv4}
\frac{dx_r}{dt} = \frac{1}{3}V_0(u)+q = \frac{2}{3}u_m,
\end{equation}
where, due to the specific initial condition $k_s=0$ at $u=u_m$ and due to the propagation along the resting medium, $u=0$ we get the second equality. Of course, the law of the soliton motion works only when the leading soliton of the DSW is almost formed, it is the asymptotic nature of the law. In this regard, we need to adapt this law to our needs of choosing an appropriate motion law of the soliton edge of the DSW. In our problem, there is a characteristic time -- the time $t1$ for the left edge of the DSW to reach the peak of the impulse. Elementary analysis and understanding of the situation show that this time is small compared to the entire decay time of the impulse $t2$, so with good accuracy, we can consider that the soliton edge of the DSW remains stationary until time $t1$. In the absence of any other speed, except for the asymptotic one, we will assume that after $t1$, the soliton edge moves with the asymptotic speed.

First, we will consider the trivial case of a soliton with a significantly higher speed than any other speeds in the problem. Obviously, in this case, we can apply a simple linear approximation for the effective background. As an example, let us consider results of numerical simulations from Section 2 with the same situations but in which we consider fast solitons. As one can see in Fig.\ref{fig4}, linear approximation expressed by dashed lines provides enough good approximation of the dynamics of fast solitons that it wasn't so in the case of slow solitons as in Fig.\ref{fig2}. The same reasoning can be applied to a pulse with finite dimensions.
\begin{figure}[t]
\begin{center}
	\includegraphics[width = 7cm,height = 7cm]{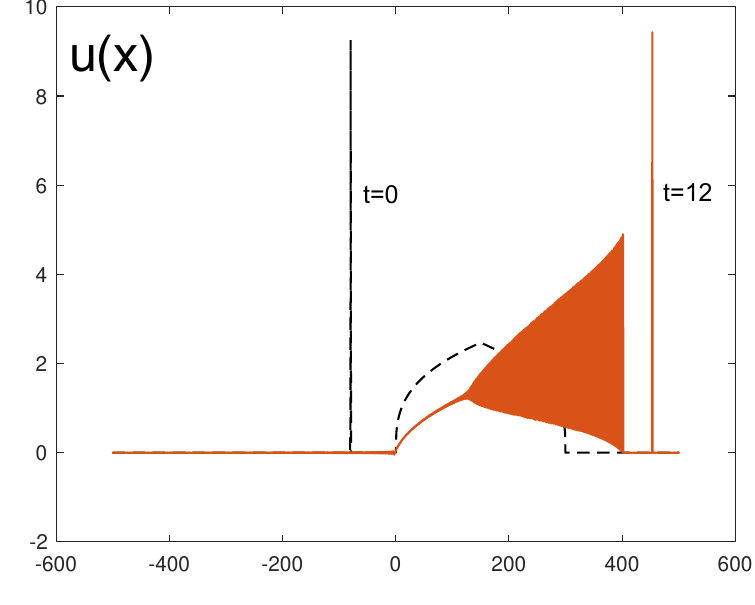}
\caption{The initial distribution \eqref{gkdv5} with $a=150, b=10$ is shown by the black dashed line at coordinate $x0=-160$ with an initial velocity $V=40$. The brown color depicts the picture at time $t=12$. The graph shows the typical size relationship between the soliton and the impulse discussed in this section. The picture is obtained numerically by using Eq.\eqref{gkdv1} with $p=3/2, \varepsilon=0.25$.
}
\label{fig5}
\end{center}
\end{figure}
The situation is quite different in the case of fast solitons, but comparable in speeds and amplitudes to the speed, for example, of the maximum of the pulse or the leading soliton (see Fig.\ref{fig5}). In this case, we need to propose some other method for selecting the effective background. To do this, let's refer to the already precisely solved problems from previous sections: it turned out that the effective field takes into account the main features and characteristics of the pulse and generally passes somewhere between the oscillations in the DSW. If we take this into account, we need to understand what can, in the most general case, characterize any pulse and ultimately pass between the oscillations. Apparently, nothing can be proposed except for the maximum amplitude $u_m$. Indirectly, many factors in the soliton science of simple waves indicate that $u_m$ plays a significant role in the evolution of the system.

Here, it should be noted that we are, of course, considering situations where the soliton interacts with a sufficiently formed DSW. If this is not the case and the soliton is near a DSW that is just beginning to form, then the propagation time will be quite short, and therefore we come to the condition for the applicability of linear approximation of the effective field.

Returning to the general equation \eqref{gkdv1}, to demonstrate the theory, we want to present here an easy-to-calculate case allowing us to do analytically a major part of the calculation without numerical techniques. So, now we are interested in the case of $p=3/2$ and an initial background wave
\begin{figure}[t]
\begin{center}
	\includegraphics[width = 7cm,height = 7cm]{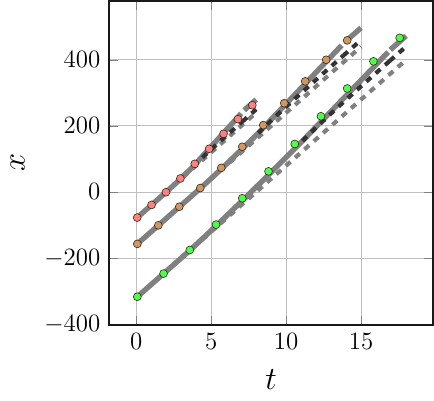}
\caption{ The soliton's pathes through the initial impulses \eqref{gkdv5} with different initial data ($V_0 = 40$) $\{a,b,x_0\}$: red $\{75,3,-80\}$, brown $\{150,10,-160\}$, green $\{250,20,-120\}$. For the last case, the graph was shifted down by $200$ units to avoid obscuring other data. The points indicate the results of numerical calculations, the gray dashed line represents motion without considering the impulse at all. The black dashed line corresponds to a linear effective field, the solid line represents the effective field as the maximum value of the initial impulse. The point of divergence between the black dashed and gray solid lines indicates the moment of the soliton's entry into the DSW. The analytical prediction of the exit point is marked by a gap in the solid line. All dependencies end both numerically and analytically at the exit from the DSW. The numerical calculation is obtained in the framework of Eq.\eqref{gkdv1} with $p=3/2$ and $\varepsilon=0.25$.
}
\label{fig6}
\end{center}
\end{figure}
\begin{equation}\label{gkdv5}
u_0(x) = 
\begin{cases}
\big(\frac{x}{b}\big)^{1/3} &\quad 0\leq x \leq a,
\\
\big(-\frac{x-2a}{b}\big)^{1/3} &\quad a\leq x,
\end{cases}
\end{equation} 
where $a,b$ are the scale coefficients (see black dashed line in Fig.\ref{fig5}). An evolution of the parts of such a background can be easily determined using the Hopf equation, it leads to 
\begin{equation}\label{gkdv6}
\big(u_{1,2}\big)^{3/2} = \frac{-3t\pm\sqrt{9t^2\pm xb}}{\pm b}, 
\end{equation} 
where index 1 corresponds to the upper sign on the right side of the equation, index 2 corresponds to the lower sign n and the replacement $x\to x-2a$. Maximum value of the pulse $u_m=u_0(a) = (a/b)^{1/3}$ moves with the velocity $6(a/b)^{1/2}$ to the right. At the initial time the right part begins to break  at $x=2a$ with the formation of DSW. The left edge expands according to \eqref{gkdv3} 
\begin{equation}\label{gkdv7}
\frac{dx_{l_1}}{dt} = 6\frac{-3t-\sqrt{9t^2-(x-2a)b}}{b} 
\end{equation}
where the solution $x_{l_1}(t) = 2a-27t^2/b$ is easily guessed. Such a law works until the edge reaches the highest point at the time that can be calculated from
\begin{equation}\label{gkdv8}
2a-27t^2/b = 6\big(a/b)^{1/2}t.
\end{equation}
After that, the left edge of the DSW starts the motion along the left part of the total background wave. The motion as well can be determined by solving 
\begin{equation}\label{gkdv9}
\frac{dx_{l_2}}{dt} = -6u(x,t)=-6\frac{-3t+\sqrt{9t^2+xb}}{b},
\end{equation}
with the initial condition $x=6(a/b)^{1/2}\cdot t_m$, where $t_m$ we denote the time of meeting from Eq.\eqref{gkdv8}. To calculate it analytically we consider the characteristic equation for the Hopf equation $x-6u^{3/2}t=\bar{x}'(u)$ (where $\bar{x}'(u)$ denotes the inverse function to the initial branch of the distribution) and differentiate it with respect to $u$, it gives
\begin{equation}\label{gkdv10}
\big(\frac{dx}{dt}-6u^{3/2}\big)\frac{dt}{du} - 9u^{1/2}t=3b u^2,
\end{equation}
where one should substitute Eq.\eqref{gkdv7}. The solution has the form
\begin{align}\label{gkdv11}
&x(u)=6u^{3/2}t(u)+u^3b,\\
&t(u) = -\frac{b}{9}u^{3/2}+\frac{u_m^{3/4}t_m+bu_m^{9/4}/9}{u^{3/4}}.
\end{align}
The right edge of DSW we assume begins to move after $t_m$ with the velocity $4u_m^{3/2}=4(a/b)^{1/2}$. 

Thus, we only need to know the law of the soliton on the left branch of the initial pulse and after on the constant value $u_m$ to describe the problem fully analytically. It can be obtained as well form Eq.\eqref{gkdv11} with soliton's velocity 
\begin{equation}\label{gkdv12}
\frac{dx}{dt} = 2 u(x,t)+q,
\end{equation}
where $q$ is a constant of the soliton. The solution of this ODE easily leads to the law of motion
\begin{align}\label{gkdv13}
&x(u)=6u^{3/2}t(u)+u^3b,\\
&t(u) = -\frac{b}{5}u^{3/2}-\frac{1}{30}bq+\frac{t_0q^{3/2}+bq^{5/2}/30}{(q-4u^{3/2})^{3/2}},
\end{align}
which is obtained with the initial condition $t(u=0)=t_0$ that means the soliton arrives at $x=0$ at $t=t_0$. According to our theory, the final stage of the propagation through DSW can be considered as a motion along the uniform field $u_m$ with a trivial law of the notion.

To numerically examine different configurations of the described problem, we will set a soliton on a zero background with a sufficiently high speed $V=40 $ not far from the initial impulse \eqref{gkdv5}; according to Eq.\eqref{gkdv12}, this speed will be a constant $q$. The form of the soliton can be easily obtained from Eq.\eqref{gkdv1} as a solution $u = u(x-V_st)$ with zero boundary conditions: 
\begin{equation}\label{gkdv13}
u_s = \frac{[V_s(p+1)(p+2)/12]^\frac{1}{p}}{\cosh^\frac{2}{p}[\frac{p\sqrt{V_s}}{2\varepsilon}(x-V_st)]},
\end{equation}
where we should put $p=3/2$ (remark: such a soliton is unstable\cite{kuzn-stability} if $p>4$). In Fig.\ref{fig6} there are three typical configuration of the problem: a narrow and high wave pulse, long and with lower amplitude, and an intermediate variant. As one can see the approximation of the uniform effective field shows a good agreement with the numerical calculation. There are slight discrepancies in the two extreme cases. In the case of the high-amplitude wave pulse, it may turn out that the soliton is not narrow enough and it senses the inhomogeneity of the pulse. In the case of the wide pulse, it may turn out that the prolonged passage through the DSW reveals the roughness we built into the theory. But in both cases, these differences are less significant against the backdrop of the unexpectedly good agreement between such a simple and rough theory and the numerical calculations.

In conclusion, we would like to warmly bid farewell to the problem of the soliton passing through the DSW after the step decay, which initiated this entire research. Recall that in article\cite{Ablow-23}, it was proposed to take a straight line as the effective field. The numerical calculation presented in this article (figure 10) demonstrate that this suggestion works well for very fast solitons, while for slow ones, the situation is significantly worse. Referring to our theory, we must conclude that for the post-step decay, the effective background should be taken equal to the maximum, that is, the amplitude of the step. This instantly leads to the fact that from the perspective of the graph, the motion of the soliton before interacting with the DSW and the motion after the interaction are almost indistinguishable. With good accuracy, this is exactly what is depicted in Figure 10 of the mentioned article. Here, it is necessary to make an important clarification: our theory works only if the amplitude of the soliton is not similar to the amplitude of the DSW. Cases like those shown in Ref.\cite{Ablow-23} (figure 18) do not conform to our proposed approach.

\section{Conclusion}

The article examined the propagation of narrow and fast solitons through a DSW. The problem was considered from the perspective of introducing an effective field instead of DSW oscillations. In the specific case of KdV and mKdV, this idea was proven using the reduction of equations expressing the law of conservation of momentum and the law of motion of the soliton mass center. In these cases, examples of the decay of power profiles were considered. It was shown that the effective field is an analogous power profile connecting the edges of the DSW. When comparing this proposal with the numerical calculation, a variant using a linear effective field was demonstrated, which showed worse results. 

The solution of these particular cases allowed us to conclude that the effective field passing through the DSW contains characteristic information about the initial form of the  background impulse. For gKdV, in the case of any finite wave impulse, the only characteristic information is the maximum amplitude, which plays an important role in many wave and soliton processes. In this regard, it was suggested that it could be taken as the effective field. This hypothesis showed good agreement with numerical calculations.

\section{Acknowledgment}

I would like to sincerely thank my teacher, A.M. Kamchatnov, who set this task and gave valuable guidance and advice. 
This research is funded by the RSF grant number 19-72-30028 (Section 4-6).

\section*{Appendix A}

Here the analytical calculation from Section 2 will be described in more detail.

We consider the soliton \eqref{kdv2} in the limit $\varepsilon \rightarrow 0$ as a delta function. In the the study of the evolution of the soliton momentum, we get the general equation \eqref{kdv7}. To calculate integrals, we use the integration by parts, taking into account the resetting of boundary terms,
\begin{equation*}
\begin{split}
&\int  vv_{xxx}dx = 0, \\
&\int  v^2v_xdx = 0, \\
&\int  vF[w]dx = F[w(x_s,t)] \int  vdx, \\
&\int v(wv)_xdx = \int \big\{\frac{w}{2} (v^2)_x + v^2w_x \big\}dx  = \frac{w_x(x_s,t)}{2} \int v^2dx,
\end{split}
\end{equation*}
The first two integrals are easily calculated using the parity of the integrand. In the last two we use the well-known property of the delta function. So, the integrals which we are now interested in are equal to
\begin{equation*}
\begin{split}
&  \int  vdx =  \frac{k^2}{2}  \int \frac{dx}{\cosh^2[\frac{k}{2\varepsilon}(x-x_s)]} = 2\varepsilon k \\
&  \int v^2dx = \frac{k^4}{4}  \int \frac{dx}{\cosh^4[\frac{k}{2\varepsilon}(x-x_s)]} = \frac{2}{3} \varepsilon k^3
\end{split}
\end{equation*}
After these preparations the general form \eqref{kdv7} can be reduced to the convenient shape \eqref{kdv9}.

The same approach we apply to the general form of the evolution of the center of the soliton mass \eqref{kdv8}:
\begin{equation*}
\begin{split}
& \int xvF[w]dx = x_sF[w(x_s,t)] \int vdx,\\
&\int xv^2v_x dx = -\int \frac{v^3}{3}dx ,\\
&\int xv(wv)_x dx = \int \{xw_x -\frac{(xw)_x}{2}  \}v^2dx = \frac{x_sw_x-w}{2}\int v^2 ,\\
&\int xvv_{xxx}dx = -\int \{vv_{xx}+xv_xv_{xx}\}dx = \int \frac{3}{2}(v_x)^2dx,\\
\end{split}
\end{equation*}  
where one should use
\begin{equation*}
\begin{split}
&  \int  v^3 dx =  \frac{k^6}{8}  \int \frac{dx}{\cosh^3[\frac{k}{2\varepsilon}(x-x_s)]} =\frac{4}{15} \varepsilon k^5 \\
&  \int (v_x)^2dx = \frac{k^5}{4\varepsilon}  \int \bigg(\frac{\sinh{\frac{k}{2\varepsilon}(x-x_s)]}}{\cosh^3[\frac{k}{2\varepsilon}(x-x_s)]}\bigg)^2 dx = \frac{2}{15\varepsilon} k^5
\end{split}
\end{equation*}
After these manipulations and reduction of the similar terms we obtain the simple shape for the evolution of the center of mass \eqref{kdv11}.

\section*{Appendix B}

Here we will demonstrate how to get the simplified equations \eqref{mkdv7} after which we arrive at the key equations \eqref{mkdv8}. Beginning with mKdV equation \eqref{mkdv1}, we have considered the soliton solution \eqref{mkdv2} and replaced the natural background $u_b$ with effective background $w$ in Eq. \eqref{mkdv3}. Then we have reduced the center mass equation for the soliton and the momentum equation to Eqs. \eqref{mkdv5},\eqref{mkdv6} using the integration by parts similarly to how it was done in Appendix A. So, now we need to take the integrals in the last two equations, where it is convenient to introduce new notation $a$, $q$ by rule
\begin{equation}\label{B1}
\begin{split}
v(x,t) &= -\frac{k^2}{2w+\sqrt{4w^2-k^2}\cosh{\big[ \frac{k}{\varepsilon} (x-x_s)}\big]} =
\\
&= - \frac{k^2}{Q}\frac{1}{a+\cosh{\big[\frac{k}{\varepsilon} (x-x_s)}\big]},
\end{split}  
\end{equation} 
that is 
\begin{equation}\label{B2}
Q^2 = 4w^2-k^2 \qquad a = \frac{2w}{Q}.
\end{equation}
One can note that $a>1$ since $Q = \sqrt{4w^2-k^2}<2w$.  Of course there are some relations between \{$a,q$\} and \{$k^2,w$\}:
\begin{equation}\label{B3}
k^2 = Q^2\cdot(a^2-1)\qquad w = \frac{aQ}{2}.
\end{equation}
The convenience of such a replacement turns out to clarify if we consider the wanted integrals $\int v^k \propto \int \frac{dx}{(a+\cosh{x})^k}$:
\begin{equation*}
\begin{split}
&\int v \propto I(a) = \int \frac{dx}{a+\cosh{x}} = \frac{2}{\sqrt{a^2-1}}\log{(a+\sqrt{a^2-1})},  \\
&\int v^2 \propto -I'(a) = \int \frac{dx}{(a+\cosh{x})^2} = \frac{aI(a)-2}{a^2-1}, \\
&\int v^3 \propto \frac{I''(a)}{2} = \int \frac{dx}{(a+\cosh{x})^3} = \frac{(2a^2+1)aI(a)-6a}{2(a^2-1)^2}, \\
&\int v^4 \propto \frac{-I'''(a)}{6} = \frac{(9+6a^2)aI(a)-8-22a^2}{6(a^2-1)^3},
\end{split}
\end{equation*}
but the most interesting case is $\int (v_x)^2$, where we will omit all non-essential multipliers:
\begin{equation}\label{B4}
\begin{split}
&\int (v_x)^2 \propto \int \frac{\big[\frac{k}{\varepsilon}\sinh{(\frac{k}{\varepsilon}x)}\big]^2+2\frac{k}{\varepsilon}a_x\sinh{(\frac{k}{\varepsilon}x)}+(a_x)^2}{\big[a+\cosh{(\frac{k}{\varepsilon}x)}\big]^4} dx
\\
&\propto \int \frac{\sinh^2{(y)}}{\big[a+\cosh{(y)}\big]^4} dy = 
\int \frac{\big[ \cosh{(y)}+a \big]^2}{\big[a+\cosh{(y)}\big]^4} dy + 
\\
&+ \int \frac{a^2-1}{\big[a+\cosh{(y)}\big]^4} dy - \int \frac{2a(\cosh{(y)}+a)}{\big[a+\cosh{(y)}\big]^4} dy = 
\\
&= -\frac{(a^2-1)}{6}I'''(a) - aI''(a)-I'(a),
\end{split}
\end{equation}
where on the first step we use a parity property and the main order of decomposition by degree $\varepsilon$, on the second step we present $\sinh^2{(y)}$ as $(\cosh^2{(y)}-1) = ( \cosh{(y)}+a )^2 - 1-a^2-2a\cosh{(y)}$ that is useful for reducing to already known integrals. Now, to simplify our calculation, let's previously assume $Q^2 = \mathrm{const}$ along the soliton motion, that is, $\frac{dQ^2}{dt}=0$, because it is exactly what we want. After all these preparations we can write Eqs. \eqref{mkdv5},\eqref{mkdv6} as
\begin{equation}\label{B5}
\begin{split}
\frac{dx_s}{dt} = &-\frac{Q^2}{(a^2-1)I'(a)}\bigg[ (a^2-1)^3I'''(a)+5a(a^2-1)I''(a)\bigg] 
\\
&- \frac{3}{2}q^2\bigg[\frac{3}{2}a^2-1\bigg],
\end{split}
\end{equation}
\begin{equation}\label{B6}
\begin{split}
3\frac{dk^2}{dt} =& \frac{2}{I'(a)}\bigg[ 6(w^2)_xk^2+2w_x\frac{k^4}{Q} - k^2\frac{da}{dt}I''(a)-
\\
&-2\overline{F}[w]QI(a) \bigg]
\end{split}
\end{equation}
To bring these formulas to a simpler form, it is necessary to make all the substitutions (it is convenient to switch to $\{k^2,w\}$ in places), then making the reverse replacement $\{Q,a\} \leftrightarrow \{k^2,w\}$, we arrive at 
\begin{equation}\label{B7}
\frac{dx_s}{dt} = -2w^2-Q^2,
\end{equation} 
\begin{equation}\label{B8}
\frac{dk^2}{dw^2} = 4,
\end{equation}
where for the second equation we can use the initial condition from \eqref{B2}. Finally, owing to additional replacement $Q^2\rightarrow q$ we obtain \eqref{mkdv7},\eqref{mkdv8}.


\begin{thebibliography}{99}

\bibitem{El-05} G. A. El, Resolution of a shock in hyperbolic systems modified by weak dispersion, Chaos, 15, 037103 (2005).

\bibitem{El-08} El, G. A., Grimshaw, R. H. J.,  Smyth, N. F., Physica D: Nonlinear Phenomena, 237(19), 2423–2435, (2008). 

\bibitem{kamch-19}A. M. Kamchatnov, Dispersive shock wave theory for nonintegrable equations, Phys. Rev. E 99, 012203 (2019).

\bibitem{El-19} T. Congy, G. A. El, M. A. Hoefer, J. Fluid Mech. 875, 1145-1174 (2019).

\bibitem{kamch-20} A. M. Kamchatnov, Theory of quasi-simple dispersive shock waves and number of solitons evolved from a nonlinear pulse, Chaos, 30, 123148 (2020).

\bibitem{kamch-shayk-21} A. M. Kamchatnov and D. V. Shaykin, Propagation of wave packets along intensive simple waves, Phys. Fluids 33, 052120 (2021).

\bibitem{kamch-shayk-23-1}D. V. Shaykin, A. M. Kamchatnov, Physics of Fluids 35, 062108 (2023).

\bibitem{kamch-23} A. M. Kamchatnov, Chaos 33, 093105 (2023)

\bibitem{kamch-shayk-23-2}A. M. Kamchatnov and D. V. Shaykin, Phys. Rev. E
108, 054205 (2023).

\bibitem{kamch-shayk-24} A. M. Kamchatnov and D. V. Shaykin, Physica D, 460, 134085 (2024).


\bibitem{Whitham-65}G. B. Whitham, Proc. R. Soc. London, Ser. A,283, 238 (1965).

\bibitem{Whitham-74} G. B. Whitham, Linear and Nonlinear Waves, Section 12, (Wiley Interscience, New York, 1974).


\bibitem{Karpman-67} V. I. Karpman, Phys. Lett. A 25, 708 (1967)


\bibitem{2018} M. D. Maiden, D. V. Anderson, N. A. Franco, G. A. El, and M. A. Hoefer, Solitonic Dispersive Hydrodynamics: Theory and Observation, Phys. Rev. Lett., 120 144101 (2018).

\bibitem{2021} K. van der Sande, G. A. El, and M. A. Hoefer, Dynamic Soliton–Mean Flow Interactionwith Non-Convex Flux, J. Fluid Mech. 928 A21 (2021).

\bibitem{Ablow-23} Mark J. Ablowitz, Justin T. Cole, Gennady A. El, Mark A. Hoefer, Xu-dan Luo, Studies in Applied Mathematics 151(3):1-62, (9 July 2023).

\bibitem{LL-M} L. D. Landau and E. M. Lifshitz, Mechanics, Nauka, Moscow 1993


\bibitem{LL-TP} Landau, L.D. and Lifshitz, E.M. (1971), The Classi- cal Theory of Fields (Volume 2 of A Course of The- oretical Physics). Pergamon Press, New York.



\bibitem{gur-pit-73} Gurevich A. V., Pitaevskii L. P., Sov. Phys. JETP 38 291, (1974); Zh. Eksp. Teor. Fiz. 65 590 (1973).


\bibitem{Stokes} G. G. Stokes, Mathematical and Physical Papers (Cambridge University Press, Cambridge, 1905), Vol. V, p. 163.


\bibitem{Tsarev} Tsarev S P Sov. Math.
Dokl. 31 488 (1985).

\bibitem{Gurevich},Gurevich A V, Krylov A L, Mazur N G Sov. Phys. JETP 68 966 (1989).


\bibitem{kuzn-stability} E. A. Kuznetsov, Soliton stability in equations of the KdV type, Phys. Lett. A 101, 314-316 (1984).


\bibitem{Kamch-Kon-04} A. M. Kamchatnov, A. Spire, and V. V. Konotop, J. Phys. A 37, 5547 (2004).


\bibitem{KamchN} A. M. Kamchatnov; Self-similar wave breaking in dispersive Korteweg-de Vries hydrodynamics. Chaos 1; 29 (2): 023106, (2019).

\bibitem{gur-pit-74} A. V. Gurevich and L. P. Pitaevskii, Sov. Phys. JETP 38, 291 (1974).

\bibitem{pot-88} G. V. Potemin, Russ. Math. Surveys 43, 39 (1988).

\bibitem{dub-93} B. A. Dubrovin and S. P. Novikov, Sov. Sci. Rev. C. Math. Phys. 9, 1 (1993).

\end{thebibliography}
\end{document}